\documentclass[prb,twocolumn,showpacs,preprintnumbers,amsmath,amssymb]{revtex4} 
\usepackage{amsmath,amssymb,amsfonts}
\usepackage{bm}
\usepackage{graphicx}
\usepackage{times}
\usepackage{color}
\usepackage[colorlinks,bookmarks=false,citecolor=blue,linkcolor=red,urlcolor=blue]{hyperref}

\begin{document}
\title{On the possible secondary component of the order parameter observed in London penetration depth measurements}
\author{A. Valli}
\affiliation{Dipartimento di Fisica, Universit\`a Sapienza, P.le Aldo Moro 2, I-00185, Roma, Italy}
\affiliation{Institute f\"{u}r Festk\"{o}rperphysik, Technische Universit\"{a}t Wien, Wiedner Hauptstrasse 8//138, 1040 Vienna, Austria}
\author{G. Sangiovanni}
\affiliation{Institute f\"{u}r Festk\"{o}rperphysik, Technische Universit\"{a}t Wien, Wiedner Hauptstrasse 8//138, 1040 Vienna, Austria}
\author{M. Capone}
\affiliation{ISC-CNR and Dipartimento di Fisica, Universit\`a Sapienza, P.le Aldo Moro 2, I-00185, Roma, Italy}
\author{C. Di Castro}
\affiliation{ISC-CNR and Dipartimento di Fisica, Universit\`a Sapienza, P.le Aldo Moro 2, I-00185, Roma, Italy}

\begin{abstract}
We discuss the effect of a secondary component of the superconducting order parameter on the superfluid density in the cuprates. If we assume a main $d_{x^2-y^2}$ gap, the most stable realization of a mixed order parameter has a time-reversal breaking $d_{x^2-y^2}+ \imath d_{xy}$ symmetry. In this state the nodes are removed and  the temperature dependence of the superfluid density changes from the linear behavior of a pure d-wave to a more rounded shape at low temperature. The latter is compatible with the  behavior experimentally observed in the in-plane magnetic field penetration depth of  optimally doped $La_{2-x}Sr_xCuO_2$ and $YBa_2Cu_3O_{7-\delta}$.


\end{abstract}
\maketitle

\section*{I. Introduction}
The identification of the pairing mechanism behind high-temperature superconductivity 
in copper oxides\cite{bednorz-muellerZPB64} remains one  of the greatest challenges in solid state physics. 
A key ingredient is the symmetry of the order parameter, 
which is expected to reflect that of the pairing interaction thus providing information on the microscopic mechanism. 
The well-established evidence of lines with vanishing amplitude in the gap function of cuprates 
along the $\Gamma$-$X$ direction of the Brillouin Zone (nodes) indicates a dominant $d_{x^2-y^2}$ symmetry 
of the order parameter \cite{dwave}, hardly compatible with the standard phonon pairing mechanism, 
which leads to an isotropic s-wave order parameter.
Anyway a small secondary component of the order parameter can develop either spontaneously or driven by external factors 
like magnetic field, doping or presence of magnetic impurities \cite{laughlinPRL80, balatskyPRL80, ramakrishnanJPCS59}. 
The development of a mixed order parameter has been also invoked to explain 
anomalies observed in the thermal conductivity in magnetic field of Bi$_2$Sr$_2$CaCu$_2$O$_8$ \cite{krishanaS277}. 
Moreover substantial deviation from the $d_{x^2-y^2}$-wave symmetry has been clearly observed 
in YBa$_2$Cu$_3$O$_{7-\delta}$ (YBCO) both in tunneling measurements \cite{deutscherEPL51}
and in laser angle-resolverd photoemission spectra revealing nodeless bulk superconductivity \cite{okawa08110479}.
A series of low-temperatures anomalies has been observed in the in-plane magnetic field penetration length 
in muon-spin rotation ($\mu$SR) experiments \cite{khasanovPRL98, khasanovJSNM, brandtPRB37, sonierRMP72}.
Experiments in  optimally doped La$_{2-x}$Sr$_x$CuO$_2$ (LSCO) and YBCO 
have indeed shown a low-temperature bump superimposed to the linear temperature behavior associated to 
d-wave superconductivity and to the presence of nodes. 
These deviations from d-wave behavior have been associated to a secondary component, which has been proposed 
to be isotropic s-wave in light of its vulnerability to a magnetic field.
Among alternative proposals, some\cite{aminPRL84, sharapovPRB73} do not assume the presence of a mixed order parameter, 
and they associate the low-temperature feature to a non-local response of the d-wave superconductor,  
which modifies the magnetic field distribution in the vortex state with respect to the standard London model.
Beside the interpretation in terms of a secondary superconducting s-wave\cite{holderEPL77}, a particle-hole 
secondary gap associated to spin density wave ordering has been invoked\cite{sharapovPRB66}.

Here we focus on the secondary superconducting gap interpretation,
and we show that only a $d_{x^2-y^2} + \imath d_{xy}$ mixed order parameter can reasonably describe the $\mu SR$ experimental results.
Previous analysis \cite{sangiovanniPRB67} has shown that, assuming a leading $d_{x^2-y^2}$ symmetry,
the most stable realization of a mixed order parameter has indeed
this time-reversal breaking $d_{x^2-y^2} + \imath d_{xy}$ symmetry.
Moreover the development of such a time-reversal breaking order parameter does not require {\it ad hoc} assumptions,
in contrast e.g. with a $d+s$ symmetry, which requires completely unrealistic parameters
as long as the $d_{x^2-y^2}$ is the dominant component of the order parameter.


This work is organized as follows: In Sec. II we present our model and approach. 
In Sec. III we discuss the general behavior of the superfluid density with a mixed order parameter and 
presents the comparison with experiments. Sec. IV contains our conclusions.


\section*{II. Model}


In this section we briefly summarize the formalism used in Ref. \onlinecite{sangiovanniPRB67} to identify 
the conditions for a secondary component to establish in the presence of a dominant $d_{x^2-y^2}$ wave.
We consider a two-dimensional square lattice characterized by the ${\cal C}_{4v}$ point group and a single band with dispersion 
\begin{equation}
 \xi_{\bf k} = -2t (\cos{\bf k}_x a - \cos{\bf k}_y a) + 4 t' \cos{\bf k}_x a \cos{\bf k}_y a - \mu ,
\end{equation}
where $t$ and $t'$ are the nearest and next-nearest hopping parameters, $\mu$ is the chemical potential 
and $a=1$ is the lattice spacing. 
Values for hopping parameters for different compounds have been chosen according to 
density-functional theory calculations in the local-density approximation \cite{andersenP4thAW}.

The aim of the present analysis is the  understanding of the competition between 
the different components of a superconducting order parameter. 
Therefore we do not attempt a solution of a microscopic model including different kind of realistic interactions, 
and we simply consider an effective low-energy interaction, 
whose strength in each symmetry channel controls the corresponding instability.
Moreover, we will study the superconducting phase within the Bardeen-Copper-Schrieffer (BCS) mean-field approach, 
which fully takes into account for the symmetry of the order parameter.
This approach is reasonably justified for instance by the relatively large doping of the samples of Refs. \onlinecite{khasanovPRL98, sonierRMP72}


We now briefly recall some relevant aspects of the BCS equations for a mixed order parameter, 
referring to \onlinecite{sangiovanniPRB67} and references therein for more details.
If we require the invariance under the symmetry of the lattice of the modulus of the order parameter, 
the latter has to transform either as an irreducible rappresentation or as a 
complex combination of the form $\Delta^{\mu} + \imath \Delta^{\nu}$ (with $\Delta^{\mu}$ and $\Delta^{\nu}$ 
transforming as two different irreducible representations) which breaks time-reversal invariance.
The development of each harmonic with a given symmetry is controlled by a specific 
spatial component of the pair potential. The isotropic s-wave is associated to the local component 
of the potential $V_0$, which is repulsive in the cuprates due to the strong Coulomb interaction. 
The $d_{x^2-y^2}$ and extended-s ($s_{x^2+y^2}$) are controlled by the nearest-neighbor coupling $V_1$, 
while the $d_{xy}$ and $s_{xy}$ (which are analogous to $d_{x^2-y^2}$ and $s_{x^2+y^2}$ with lobes along 
the diagonal directions in the plane)  are related to the next-neighbor coupling $V_2$. 
Here we will simply assume that $V_0$ is repulsive and that $V_1$ and $V_2$ are attractive. 
For the sake of definiteness we report the equations for the $d_{x^2-y^2}+ \imath d_{xy}$ mixed order parameter
\begin{equation} \label{eq:bcs}
 \left \{
  \begin{array}{lll}
   \displaystyle{\frac{1}{V_1}} &=& - \sum_{\bf k} \omega^2_d({\bf k}) \displaystyle{ \frac{1}{ 2 \epsilon_{\bf k} } \tanh \Big( \frac{1}{2} \beta \epsilon_{\bf k} \Big) }\\
 \\
   \displaystyle{\frac{1}{V_2}} &=& - \sum_{\bf k} \omega^2_{d^{\prime}}({\bf k}) \displaystyle{ \frac{1}{ 2 \epsilon_{\bf k} } \tanh \Big( \frac{1}{2} \beta \epsilon_{\bf k} \Big) }\\
\\
   \displaystyle{n} &=& 1 - \sum_{\bf k} \displaystyle{ \frac{\xi_{\bf k}}{\epsilon_{\bf k}} \tanh \Big( \frac{1}{2} \beta \epsilon_{\bf k}\Big) }
  \end{array}
 \right.
\end{equation}
Here $\beta = 1/T$ is the inverse temperature, 
$\omega_{d}({\bf{k}}) = cos({\bf k}_x a)-cos({\bf k}_y a)$ and $\omega_{d^{\prime}}({\bf k}) = 2sin({\bf k}_x a)sin({\bf k}_y a)$ 
are the harmonics associated to $d_{x^2-y^2}$ and $d_{xy}$-wave respectively, 
$\Delta_d$ and $\Delta_{d^{\prime}}$ are the associated components of the gap and
$\epsilon_{\bf k} = \sqrt{\xi^2_{\bf k} + \Delta^2_{d} \omega^2_{d}({\bf k}) + \Delta^2_{d'} \omega^2_{d'}({\bf k}) }$.
$\Delta_d$, $\Delta_{d'}$ and the chemical potential are derived solving self-consistently Eqs. (\ref{eq:bcs}). 
An energy cutoff $\omega_0$ is used in the first two k-sums.

For realistic dispersions, the $d_{x^2-y^2}$ symmetry is the leading instability 
for small dopings due to the Van Hove singularity (VHS)\cite{vanhovePR89}. 
When the main  $d_{x^2-y^2}$ order parameter appears  at $T_c$, for $T < T_c$ 
the effective dispersion $\epsilon_{{\bf k}}$ is gapped and any secondary instability requires a minimum (critical) value 
for the associated interaction strength, as opposed to the case of an instability developing in a Fermi sea ground state. 
The $d_{xy}$ component turns out to be the best candidate for the secondary gap 
(i.e., it has the lowest critical value of the interaction) since it has the largest contributions 
from the regions in which the main gap has nodes. Since the onset of a secondary component is essentially determined 
by the competition with the main gap, one can favor a mixed state by reducing the $d_{x^2-y^2}$ component. 
The complementarity between $d_{xy}$ and $d_{x^2-y^2}$ also implies that the two components 
can exist symultaneously for a wide range of parameters. 
Other instability channels require much larger couplings and, even more importantly, 
hardly give rise to a ``coexistence" of order parameters. 
In most cases, and in particular for s-wave components, the secondary order parameter 
can less efficiently exploit the Fermi-surface portions in which the first gap has nodes. 
Therefore, if we increase the associated coupling, we have an abrupt change from a pure $d_{x^2-y^2}$ to a pure 
order parameter of different symmetry, and a very fine tuning is required to have both order parameters. 

The focus of this paper is the effect of a secondary component of the superconducting order parameter 
on the superfluid density $\rho_s$, which is directly related to the London penetration depth by the relation 
$\lambda^{-2}= 4\pi e^2 \rho_s / mc^2$, being $m$ the electron mass 
and $c$ the speed of light. $\rho_s$ is defined as
\begin{equation}
 \rho_s = \sum_{\sigma} \frac{\partial^2\xi_{\bf k}}{\partial {\bf k}^2} \langle c^{\dag}_{{\bf k} \sigma} c_{{\bf k} \sigma} \rangle - \lim_{k \rightarrow 0} \int_{0}^{\beta} d\tau \langle j({\bf k} \tau) j({\bf -k} 0) \rangle ,
\end{equation}
where the first term is the zero-temperature contribution, while the other is the current-current response.
For BCS pairing, in case of spin degeneracy, the previous expression then reads
\begin{equation}
 \rho_s = \sum_{\bf k} \frac{\partial^2\xi_{\bf k}}{\partial {\bf k}^2} \bigg[ 1 - \frac{\xi_{\bf k}}{\epsilon_{\bf k}} \tanh \Big( \frac{\beta \epsilon_{\bf k}}{2} \Big) \bigg] + 2 \sum_{\bf k} \Big( \frac{\partial \xi_{\bf k}}{\partial {\bf k}} \Big)^2 \frac{\partial f(\epsilon_{\bf k})}{\partial \epsilon_{\bf k}} ,
\end{equation}
being $f(\epsilon_{\bf k}) = 1/(e^{\beta \epsilon_{\bf k}} + 1)$ the Fermi distribution function for the Bogoljiubov quasiparticles.

\section*{III. Superfluid Density in the mixed state}

\begin{widetext}
 
\begin{figure}[ht!]
\begin{center}
\includegraphics[width=0.90\columnwidth]{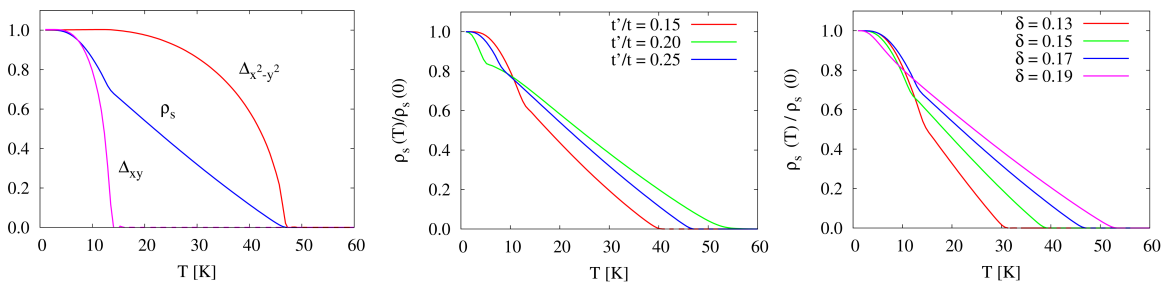}
\vspace{-0.3cm}
\caption{\small (a) $\rho_s$ and superconducting gaps $\Delta_{x^2-y^2}$ and $\Delta_{xy}$ as a function of $T$. 
Each quantity is normalized to its $T=0$ value in order to confront the curves: it is clear as the low-temperature feature 
of the superfluid density relies on the existence of a secondary component of the order parameter. 
(b) Behavior of $\rho_s$ for different values of $t'/t$ for $\delta = 0.17$; 
(c) Behavior of $\rho_s$ for different dopings at fixed $t'/t =0.15$}
\label{fig:rho}
\end{center}
\end{figure}



\end{widetext}

Before addressing the comparison with experimental data, we consider the effect of the onset 
of the $d_{x^2-y^2}+id_{xy}$ mixed order parameter in general terms. 
In Fig. \ref{fig:rho}a we plot the temperature dependence of the superfluid density 
and of the components of the superconducting gap (all normalized to their $T=0$ values) for 
$V_1/t=0.50$, $V_2/t=1.05$, $\omega_0/t=0.25$ and $t'/t=0.25$\cite{footnote1}.
Indeed the results show that the linear temperature behavior characteristic of the 
d-wave state\cite{sharapovPRB73}, associated to the presence of nodal quasi particles, 
is modified below a temperature $T'_{c}$ (see Fig. \ref{fig:rho}a). 
The low-temperature feature of the superfluid density is clearly related to the development of a $d_{xy}$ gap, 
which fills the nodes of the $d_{x^2-y^2}$ component below the secondary ``critical temperature'' $T'_c$. 
In this regime, in which a $d_{x^2-y^2} + \imath d_{xy}$ order parameter is stable, 
the shape of the superfluid density is more similar to that of a usual $s-$wave superconductor, 
reflecting the absence of low energy excitations. 
This shows that an ``s-wave-like" behavior at low temperatures does not automatically suggest an s-wave component, 
and that the $d_{x^2-y^2}+id_{xy}$ order parameter generates a temperature behavior 
which reproduces the qualitative results of Refs. \onlinecite{khasanovPRL98,khasanovJSNM}.

We now briefly discuss  how the shape of the superfluid density depends the parameters of the system. 
A crucial parameter which varies in the different cuprates is the next-neighbor hopping $t^{\prime}$ \cite{andersenP4thAW}
which controls the position of the VHS. Therefore $t'$ can push the singularity close to the chemical potential, 
thereby favoring the $d_{x^2-y^2}$ at the expenses of the secondary gap. 
Indeed, as shown in Fig.  \ref{fig:rho}b) at fixed doping $T'_c$  decreases as the chemical potential approaches the VHS. 
The same behavior holds for the amplitude of the secondary gap as expected within BCS.
Similar results are naturally obtained by changing the hole concentration instead of $t'$, 
i.e. shifting the chemical potential and preventing it to lie within the cutoff energy range from the VHS. 
In practice the variation of $t'$ in different materials can be quite large, 
and it affects the symmetry of the order parameter much more than the doping, 
if the latter is taken in the physically relevant regime.
In some cases (e.g. in LSCO compounds, where $t'/t \simeq 0.15$ \cite{andersenP4thAW}), the chemical potential 
can approach or cross the VHS in the relevant doping range (see Fig. \ref{fig:rho}c). 
On the other hand when $t'$ is larger (e.g. in YBCO compounds) and the singularity is far from the Fermi level, 
the effect of doping becomes less important.
Also orthorombic distortions or bilayer splitting reduce the $d_{x^2-y^2}$ gap, 
allowing for a larger secondary component\cite{sangiovanniPRB67}. 
As a more technical note, the value of the cutoff $\omega_0$ plays a role in the stability of the secondary component 
because it selects the portion of density of states which contributes to the effective coupling, i.e. a small cutoff makes 
the system more sensitive to the details of the bandstructure. 
For the range of parameters of interest this reflects in a stronger effect of the VHS, 
which favors the main component at the expenses of the secondary one. 

\begin{figure}[ht!]
\begin{center}
\includegraphics[width=0.80\columnwidth]{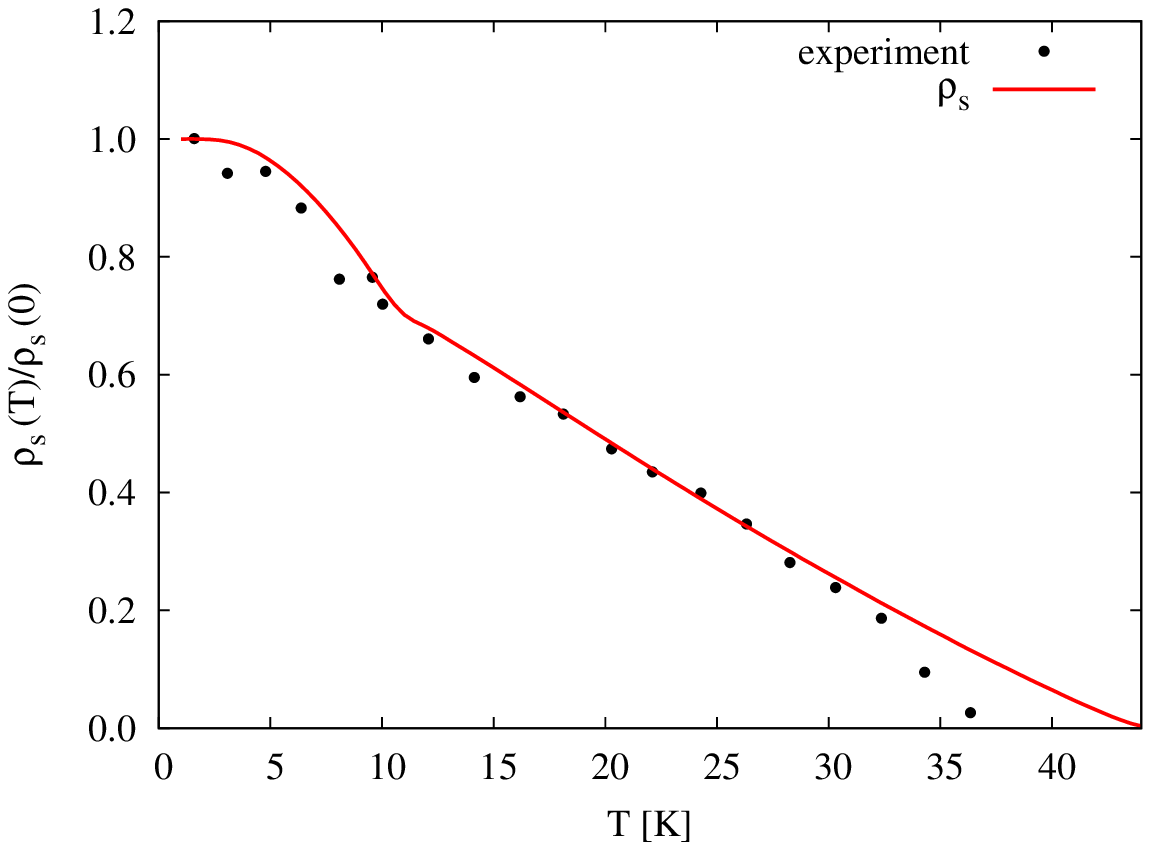}
\includegraphics[width=0.80\columnwidth]{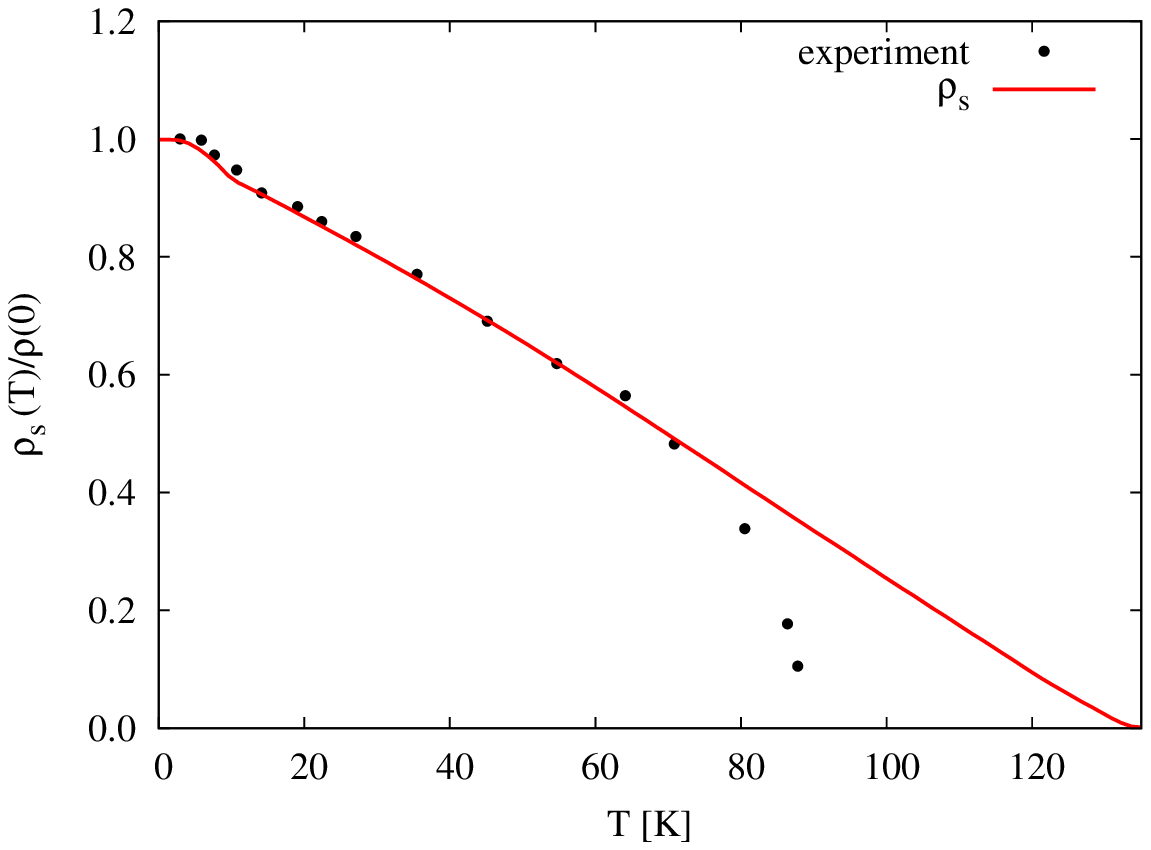}
\vspace{-0.3cm}
\caption{\small Theoretical results for superfluid density against experimental data 
for the (normalized) spin depolarization rate on $La_{1.85}Sr_{0.15}CuO_4$ \cite{khasanovPRL98} 
and $YBa_2Cu_3O_{6.95}$ \cite{sonierRMP72} compounds. }
\label{fig:exp}
\end{center}
\end{figure}

We now turn to the experimental evidences discussed above considering the specific cases 
of optimally doped LSCO\cite{khasanovPRL98} and  YBCO \cite{sonierRMP72}. 
We use parameters ($V_1/t = 0.55$, $V_2/t = 1.1$, $\omega_0/t=0.25$, $t'/t=0.135$ for LSCO 
and $V_1/t = 1.1$, $V_2/t = 1.275$, $\omega_0/t=0.25$, $t'/t=0.35$ for YBCO) 
that reproduce the experimental dispersions and the zero-temperature value of the gaps. 
The doping is $\delta = 0.17$ in both cases. As shown in Fig. \ref{fig:exp}, our simple theoretical approach  
well reproduces the temperature behavior of $\rho_s$ for a wide range of temperature. 
The appearance of the secondary component is much more pronounced for LSCO, 
in agreement with the above analysis about the role of $t'/t$. 
The deviation between the BCS results and the experiments close to $T_c$ are obviously expected 
because of the relevance of fluctuations for quasi two-dimensional strong-coupling superconductors.

Our analysis shows that the experimental evidence of a low-temperature ``bump" on top of the linear temperature dependence 
can be understood in terms of a $d_{x^2-y^2}+id_{xy}$ order parameter without invoking an s-wave component. 
On the other hand an isotropic component is extremely hard to establish 
when the main order parameter has $d_{x^2-y^2}$ symmetry \cite{sangiovanniPRB67}. 
In particular an s-wave order parameter requires a huge local attractive interaction, in clear contradiction 
with the almost universally recognized role of Coulomb repulsion. We emphasize that even if we accept an attractive 
local component, the $s$-wave and the $d_{x^2-y^2}$-wave gaps hardly coexist in the same Fermi surface. 
It is therefore almost impossible to reproduce the experimental behavior with an isotropic secondary gap 
without making totally {\it{ad hoc}} and unrealistic assumptions  and a very fine tuning of parameters         .

As a final remark we focus our attention on the effect of an external magnetic field, which seems to flatten out 
the low-temperature  behavior of the penetration depth in the experimental data\cite{khasanovPRL98,khasanovJSNM}. 
Several conflicting interpretations have been proposed. 
Some of them \cite{aminPRL84, sharapovPRB73} do not rely on the presence of a mixed order parameter, 
but relate the flattening to a non-local response of the d-wave superconductor, 
which modifies the magnetic field distribution in the vortex lattice with respect to the standard London model.
Other studies identify the low-temperature feature with a second gap being either spin density wave\cite{sharapovPRB66} 
or a different superconducting gap in the same spirit of the present analysis \cite{holderEPL77}.
In Ref \onlinecite{holderEPL77} in particular, the fragility of the secondary component to an external magnetic field 
has been advocated as a proof of its '$s$-wave nature. 
It is crucial to observe that the experimental magnetic fields (0.02, 0.1, 0.64 T)\cite{khasanovPRL98} 
are too low to directly affect the secondary gap, whatever the mechanism could be. 
In this sense, the effect of the field can only be a minor indirect consequence, 
and it hardly shed lights on the symmetry of the secondary component.
Within the second gap interpretation, we notice that the main element for the stability of the secondary component 
is actually the size of the main gap, and that tiny variations of the latter may completely suppress the former.

\section*{IV. Conclusions}

In this paper we have shown the effect of a time-reversal breaking order parameter $d_{x^2-y^2}+id_{xy}$ on 
the temperature evolution of the superfluid density within a BCS formalism. As previously shown, this combination is the most stable mixed order parameter if the main component has $d_{x^2-y^2}$ symmetry. Moreover, it is essentially the only way to have a smooth evolution from a pure d-wave to a superconducting phase which displays a secondary component at low temperature.  The same smooth evolution is mirrored in the temperature behavior of the superfluid density, in which a small ``bump" is superimposed to the linear behavior characteristic of a pure  $d_{x^2-y^2}$-wave.
	
We compared numerical results to experimental data on two cuprates and showed that the low-temperature feature 
observed in $\mu$SR measurements can be reproduced assuming reasonable parameters for the 
system in such an unconventional symmetry phase.

\section*{Acknowledgments}
We acknowledge L. Benfatto and K. Held for useful discussions and 
 financial support of MIUR PRIN 2007 Plot. 2007DW3MJX003 and FWF Science College WK004.
M.C. is funded by FP7 ERC Starting Independent Research Grant ``SUPERBAD" 
(Grant Agreement n. 240524).


\begin{thebibliography}{99}
\bibitem{footnote1} one could get similar results with $V_1>V_2$ using a significantly samller cutoff which makes the numerical evaluation of the k-sums computationally heavier.
\end{thebibliography}
\end{document}